\documentclass[aps,twocolumn,groupedaddress]{revtex4-1}

\usepackage{graphicx}
\usepackage{bm}
\usepackage{tabularx}
\usepackage{amsmath}

\begin{document}

\preprint{}

\title{High-$T_c$ superconductivity and antiferromagnetism in multilayer cuprates: \\$^{63}$Cu- and $^{19}$F-NMR on five-layer Ba$_2$Ca$_4$Cu$_5$O$_{10}$(F,O)$_2$}

\author{Sunao Shimizu}
\email[]{E-mail: sshimizu@riken.jp}
\altaffiliation[]{Present address: Correlated Electron Research Group (CERG), RIKEN Advanced Science Institute (ASI), Wako, Saitama 351-0198, Japan}

\author{Shin-ichiro Tabata}
\author{Shiho Iwai}
\author{Hidekazu Mukuda}
\author{Yoshio Kitaoka}
\affiliation{Graduate School of Engineering Science, Osaka University, Toyonaka, Osaka 560-8531, Japan }
\author{Parasharam M. Shirage}
\author{Hijiri Kito}
\author{Akira Iyo}
\affiliation{National Institute of Advanced Industrial Science and Technology (AIST), Umezono, Tsukuba 305-8568, Japan}

\date{\today}

\begin{abstract}
We report systematic Cu- and F-NMR measurements of five-layered high-$T_c$ cuprates Ba$_2$Ca$_4$Cu$_5$O$_{10}$(F,O)$_2$. It is revealed that antiferromagnetism (AFM) uniformly coexists with superconductivity (SC) in underdoped regions, and that the critical hole density $p_c$ for AFM is $\sim$ 0.11 in the five-layered compound. We present the layer-number dependence of AFM and SC phase diagrams in hole-doped cuprates, where $p_c$ for $n$-layered compounds, $p_c(n)$, increases from $p_c(1)$ $\sim$ 0.02 in LSCO or $p_c(2)$ $\sim$ 0.05 in YBCO to $p_c(5)$ $\sim$ 0.11. The variation of $p_c(n)$ is attributed to interlayer magnetic coupling, which becomes stronger with increasing $n$.
In addition, we focus on the ground-state phase diagram of CuO$_2$ planes, where AFM metallic states in slightly doped Mott insulators change into the uniformly mixed phase of AFM and SC and into simple $d$-wave SC states. The maximum $T_c$ exists just outside the quantum critical hole density, at which AFM moments on a CuO$_2$ plane collapse at the ground state, indicating an intimate relationship between AFM and SC. 
These characteristics of the ground state are accounted for by the {\it Mott physics} based on the $t$-$J$ model; the attractive interaction of high-$T_c$ SC, which raises $T_c$ as high as 160 K, is an in-plane superexchange interaction $J_{in}$ ($\sim 0.12$ eV), and the large $J_{in}$ binds electrons of opposite spins between neighboring sites. It is the Coulomb repulsive interaction $U$~($ > 6$ eV) between Cu-3$d$ electrons that plays a central role in the physics behind high-$T_c$ phenomena.

\end{abstract}


\maketitle

\section{Introduction}

Despite extensive research for more than a quarter of a century, there is still no universally accepted theory about the mechanism of superconductivity (SC) in copper-oxide superconductors. The main controversy exists over the attractive force that forms Cooper pairs, which leads to a remarkable high SC transition temperature $T_c$.
It has been known that the hybridization between Cu($3d_{x^2-y^2}$) and O($2p\sigma$) orbitals induces a large in-plane superexchange interaction $J_{in}\sim 0.12$ eV or 1300 K in nondoped CuO$_2$ planes, and that strong antiferromagnetic (AFM) correlations exist even in SC states. Experiments have observed local magnetic moments in metallic or SC regions \cite{Weidinger,Tranquada,Niedermayer,Lee,Sidis,Lake,Sanna,Miller,Das,JulienLSCO,Stock,Haug,Coneri}, and theories have pointed out a strong relationship between AFM and SC in underdoped regions \cite{Chen,Giamarchi,Inaba,Anderson1,Zhang,Himeda,Kotliar,TKLee,Demler,Shih,Yamase,Paramekanti,Anderson2,Senechal,Capone,Ogata,Pathak,Watanabe}. 
Those reports have shown that AFM plays a key role in the mechanism of high-$T_c$ SC in cuprates.

We have reported on AFM and SC properties in multilayered cuprates \cite{Mukuda,Shimizu0234F,ShimizuFIN,Shimizu0223F}, where long-range static AFM orders uniformly coexist with SC in underdoped regions \cite{ShimizuFIN}. 
On the other hand, the uniform coexistence has not been observed in typical high-$T_c$ cuprates such as single-layered La$_{2-x}$Sr$_x$CuO$_4$ (LSCO) \cite{JulienLSCO,Keimer} and bi-layered YBa$_2$Cu$_3$O$_{6+y}$ (YBCO) \cite{Sanna,Coneri}.
The difference of the phase diagrams in various cuprates is possibly attributed to the strength of interlayer magnetic coupling, which would become stronger with an increase in the stacking number $n$ of CuO$_2$ layers in a unit cell. In addition to that, it is expected that disorder effects, which is in association with chemical substitutions, are enhanced when $n$ is small \cite{JulienLSCO,Coneri,Albenque,Alloul}, and that such a disorder masks the intrinsic nature of CuO$_2$ planes.
In order to understand the relationship between AFM and SC, therefore, it is necessary to investigate the $n$ dependence of electronic properties in underdoped regions.

\begin{figure*}[t]
\begin{center}
\includegraphics[width=0.8\linewidth]{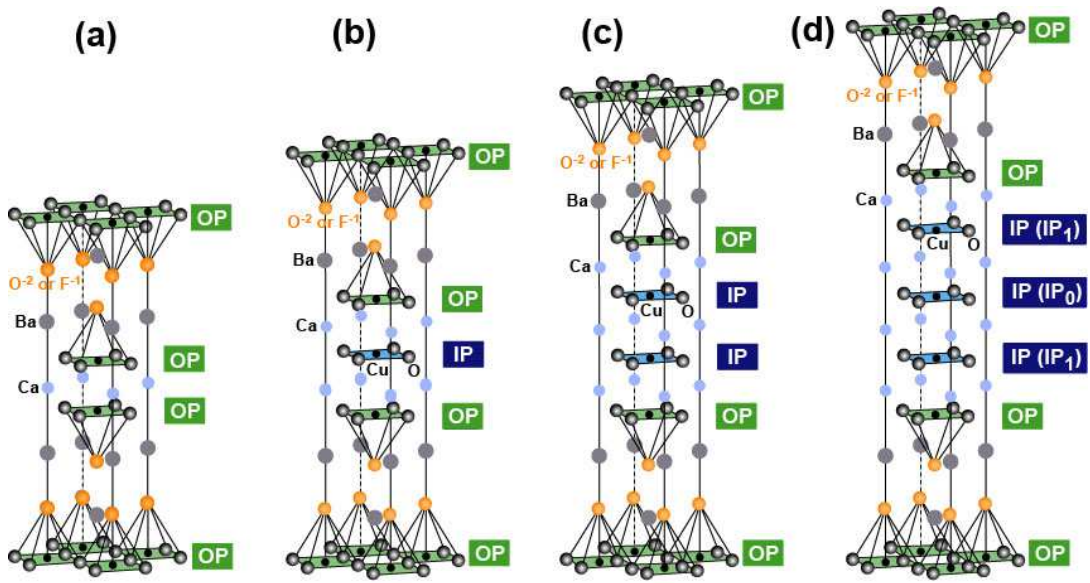}
\end{center}
\caption{ \footnotesize  (color online) Crystal structures of Ba$_2$Ca$_{n-1}$Cu$_n$O$_{2n}$(F,O)$_2$ with (a)  $n$=2 (0212F), (b) $n$=3 (0223F), (c) $n$=4 (0234F), and (d) $n$=5 (0245F). Here, $n$ is the number of stacked CuO$_2$ layers. 
The heterovalent substitution of F$^{1-}$ for O$^{2-}$ at apical sites decreases the hole density $p$. Outer and inner CuO$_2$ planes are denoted as OP and IP, respectively.
Note that, in a precise sense, there are two kinds of IP layers: the innermost inner CuO$_2$ plane (IP$_0$) and the two adjacent inner ones (IP$_1$).
}
\label{fig:crystal}
\end{figure*}

Apical-fluorine multilayered high-$T_c$ cuprates Ba$_2$Ca$_{n-1}$Cu$_n$O$_{2n}$(F,O)$_2$ (02(n-1)nF) provide us with the opportunity to investigate the $n$-dependent AFM-SC phase diagrams. 02(n-1)nF comprises a stack of $n$ CuO$_2$ layers, as shown in Figs.~\ref{fig:crystal}(a)-\ref{fig:crystal}(d), and consists of inequivalent CuO$_2$ layers: an outer CuO$_2$ plane (OP) in a five-fold pyramidal coordination and an inner CuO$_2$ plane (IP) in a four-fold square coordination.
The substitution of oxygen O$^{2-}$ for apical fluorine F$^{1-}$ results in doping hole carriers, increasing $T_c$ from underdoped to optimally-doped regions \cite{Iyo2,Shirage,Shimizu0234F,Shimizu0223F,ShimizuP}. 

In this paper, we report the AFM and SC phase diagram in five-layered Ba$_2$Ca$_4$Cu$_5$O$_{10}$(F,O)$_2$ (0245F) by means of $^{63}$Cu- and $^{19}$F-NMR, and present the $n$-dependence of the phase diagram from $n$=1 to 5. 
We highlight the fact that the ground-state phase diagram derived from the present study is in good agreement with theoretical predictions based on the $t$-$J$ model. This result supports the idea that the in-plane superexchange interaction $J_{in}$ plays a vital role as the glue to form Cooper pairs or mobile spin-singlet pairs. 

The contents of this paper are as follows: after experimental details in Section II, we provide experimental results and discussions; in Section III, we show systematic Cu- and F-NMR measurements on four 0245F samples, which is evidence of the uniform coexistence of AFM and SC; in Section IV, we present the $n$-dependence of the AFM and SC phase diagrams in 02(n-1)nF, and finally construct the ground-state phase diagram inherent in hole-doped cuprates.

\section{Experimental details}

Polycrystalline powder samples used in this study were prepared by high-pressure synthesis, as described elsewhere~\cite{Iyo1,Iyo2}. Powder X-ray diffraction analysis shows that the samples comprise almost a single phase. As discussed later, the sharp Cu-NMR spectral width at $T$=300 K assures the good quality of the samples. We prepared four samples of Ba$_2$Ca$_4$Cu$_5$O$_{10}$(F$_y$O$_{1-y}$)$_2$ (0245F) and determined the values of $T_c$ by the onset of SC diamagnetism using a dc SQUID magnetometer. 
The four 0245F samples exhibit a systematic decrease in $T_c$, as listed in Table \ref{t:ggg}, as the nominal amount of fluorine F$^{1-}$ (i.e., $y$) increases.
The heterovalent substitution of F$^{1-}$ for O$^{2-}$ at apical sites (see Fig.~\ref{fig:crystal}) decreases the hole density $p$ in apical-F compounds \cite{Iyo2,Shimizu0234F,Shimizu0223F,Shirage,ShimizuP}. Note that it is difficult to investigate the actual fraction of F$^{1-}$ and O$^{2-}$ \cite{Shirage,ShimizuPRB}. 
For NMR measurements, the powder samples were aligned along the $c$ axis in an external field $H_{ex}$ of $\sim$ 16 T and fixed using stycast 1266 epoxy. NMR experiments were performed by a conventional spin-echo method in the temperature ($T$) range of 1.5 $-$ 300 K.

\begin{table*}[t]
\caption[]{List of physical properties in Ba$_2$Ca$_4$Cu$_5$O$_{10}$(F$_y$O$_{1-y}$)$_2$ (0245F) used in this study. The $T_c$ values were determined by the onset of SC diamagnetism using a dc SQUID magnetometer. Here, note that the amount of F$^{1-}$, $y$, is nominal. It is difficult to determine the actual fraction of F$^{-1}$ and O$^{2-}$ at the apical sites \cite{Shirage,ShimizuPRB}. The hole density $p$, the nuclear quadrupole frequency $^{63}\nu_Q$, and the full-width at half-maximum (FWHM) are separately evaluated for OP and IP from the Cu-NMR measurements at $T$=300 K (see text).}
\begin{center}
{\tabcolsep = 3mm
 \renewcommand\arraystretch{1.5}
  \begin{tabular}{c|cccccccc}
    \hline\hline
     0245F       & $T_c$  & $y$ & $p$(OP) & $p$(IP)  & $^{63}\nu_Q$(OP) & $^{63}\nu_Q$(IP) & FWHM(OP) & FWHM(IP)  \\
    \hline 
 $\sharp$1 &   85 K & 0.6  &0.126   & 0.066   &   13.6 MHz       & 8.3 MHz          & $\sim$ 300 Oe  & $\sim$ 110 Oe   \\
 $\sharp$2 &   75 K & 0.7  &0.106   & 0.062   &   13.1 MHz       & 7.8 MHz          & $\sim$ 230 Oe  & $\sim$  60 Oe   \\
 $\sharp$3 &   65 K & 0.8  &0.083   & 0.053   &   11.9 MHz       &    --            & $\sim$ 210 Oe  &   --      \\
 $\sharp$4 &   52 K & 1.0  &0.060   & 0.046   &   11.6 MHz       &    --            & $\sim$ 190 Oe  &   --      \\
    \hline\hline
    \end{tabular}}
 \end{center}
\label{t:ggg}
\end{table*}

\section{Results}

\subsection{Cu-NMR}

\subsubsection{$^{63}$Cu-NMR spectra and estimation of nuclear quadrupole frequency $^{63}\nu_Q$}

\begin{figure}[htpb]
\begin{center}
\includegraphics[width=1.0\linewidth]{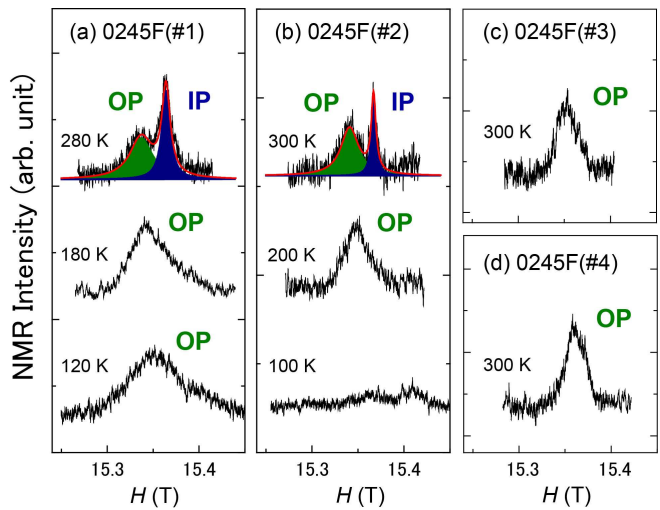}
\end{center}
\caption{\footnotesize (color online)  Typical $^{63}$Cu-NMR spectra for (a) 0245F($\sharp$1), (b) 0245F($\sharp$2), (c) 0245F($\sharp$3), and (d) 0245F($\sharp$4).
The spectra were measured with $H_{ex}\perp$ $c$ and at $\omega_0$=174.2 MHz.
For (a) 0245F($\sharp$1) and (b) 0245F($\sharp$2), the spectra of IP disappear at low temperatures due to the development of antiferromagnetic correlations. For (c) 0245F($\sharp$3) and (d) 0245F($\sharp$4), no signal is obtained for IP even at $T$=300 K.}
\label{fig:NMR}
\end{figure}

Figures \ref{fig:NMR}(a), \ref{fig:NMR}(b), \ref{fig:NMR}(c), and \ref{fig:NMR}(d) show typical $^{63}$Cu-NMR spectra of the central transition (1/2 $\Leftrightarrow$ $-$1/2) for 0245F($\sharp$1), 0245F($\sharp$2), 0245F($\sharp$3), and 0245F($\sharp$4), respectively. The field-swept NMR spectra were measured with $H_{ex}$ perpendicular to the $c$ axis ($H_{ex}\perp c$), and the NMR frequency $\omega_0$ was fixed at 174.2 MHz. As shown in Fig.~\ref{fig:crystal}(d), 0245F has two kinds of CuO$_2$ planes: an outer CuO$_2$ plane (OP) and an inner plane (IP). Therefore, the two peaks in the NMR spectra correspond to OP and IP. The assignment of NMR spectra to OP and IP has been already reported in previous literature \cite{Julien,ZhengTl2223,Kotegawa2001}. Here, note that Cu-NMR spectra for the innermost  IP (IP$_0$ in Fig. \ref{fig:crystal}(d)) and the two adjacent IP (IP$_1$ in Fig. \ref{fig:crystal}(d)) overlap each other, which suggests that their local doping levels are not so much different \cite{ShimizuP}.  Henceforth, we do not distinguish IP$_0$ and IP$_1$ in this paper. 
In 0245F($\sharp$1), the Cu-NMR spectra for both OP and IP are observed at $T$=280 K, whereas the IP's spectrum disappears at low temperatures, as shown in Fig.~\ref{fig:NMR}(a). This is because AFM correlations develop upon cooling as in the case of three-layered 0223F \cite{Shimizu0223F} and four-layered 0234F \cite{Shimizu0234F}.
We have also reported on the loss of Cu-NMR intensity due to spin dynamics in a previous paper \cite{Mukuda}. 
 In 0245F($\sharp$2), the Cu-NMR spectra are observed for OP and IP at $T$=300 K, as shown in Fig.~\ref{fig:NMR}(b). When decreasing $T$, however, not only the spectrum of IP but also that of OP disappears at low temperatures. Moreover, in 0245F($\sharp$3) and 0245F($\sharp$4), the IP's spectra are not observed even at $T$=300 K. These marked differences in the NMR spectra among the four samples suggest that AFM correlations become stronger as $p$ decreases from 0245F($\sharp$1) to 0245F($\sharp$4). The values of $p$ for the present samples, listed in Table \ref{t:ggg}, are discussed later. Here, note that the full-width at the half maximum (FWHM) of the NMR spectra at $T$=300 K decreases from 0245F($\sharp$1) to 0245F($\sharp$4), as presented in Table~\ref{t:ggg}. 
This is because the disorder associated with the atomic substitution is reduced with increasing $y$, i.e., decreasing the amount of the O$^{-2}$ substitution at the apical-F sites. 
In the same sample, the NMR spectral width for OP is much broader than that for IP; OP is closer to the Ba-F layer (see Fig. \ref{fig:crystal}), which is the source of the disorder due to the atomic substitution at apical-F sites. The values of FWHM for OP are less than $\sim$ 300 Oe, which is narrower than those for Bi- and Tl-compounds \cite{Ishida,Magishi}. On the other hand, FWHM for IP is less than $\sim$ 110 Oe, which is comparable to or even narrower than those for Y- and Hg-compounds \cite{Itoh}. These NMR linewidths point to the good quality of the present samples.

\begin{figure}[htpb]
\begin{center}
\includegraphics[width=0.9\linewidth]{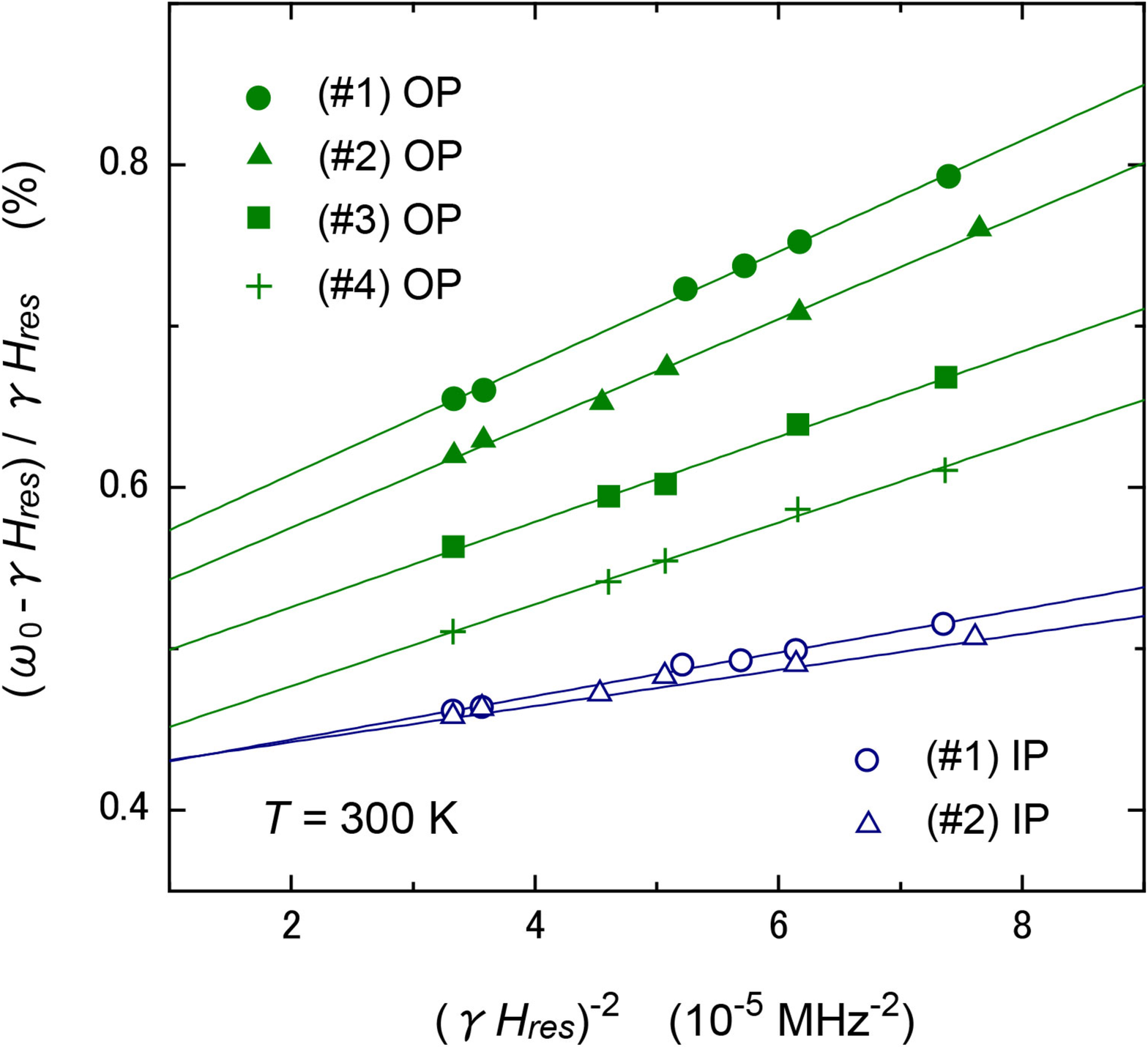}
\end{center}
\caption{\footnotesize (color online)  NMR frequency $\omega_0$-dependence of $H_{res}$ plotted as ($\omega_0 - \gamma_N H_{res}$)/$\gamma_N H_{res}$ vs ($\gamma_N H_{res}$)$^{-2}$. According to Eq.~(\ref{eq:shift}), the values of $^{63}\nu_Q$ are estimated as listed in Table~\ref{t:ggg}. }
\label{fig:nu_Q}
\end{figure}

According to the second-order perturbation theory for the nuclear Hamiltonian with $H_{ex}\perp c$ \cite{Abragam,TakigawaNQRshift}, the NMR shifts of the spectra in Fig. \ref{fig:NMR} consist of the Knight shift $K$ and the second-order quadrupole shift. The NMR shifts are expressed as 
\begin{equation}
\frac{\omega_0 - \gamma_N H_{res}}{\gamma_N H_{res}} = K+\frac{3\nu_Q^2}{16(1+K)}\frac{1}{(\gamma_N H_{res})^2}~, 
\label{eq:shift}
\end{equation}
where $\gamma_N$ is a nuclear gyromagnetic ratio, $H_{res}$ is an NMR resonance field, and $\nu_Q$ is a nuclear quadrupole frequency. 
In order to estimate $^{63}\nu_Q$ for the present samples, we have measured the $\omega_0$ dependence of $H_{res}$ at $T$=300 K in a range of 110.2 - 176.2 MHz. The obtained data set of $\omega_0$ and $H_{res}$ is plotted as ($\omega_0 - \gamma_N H_{res}$)/$\gamma_N H_{res}$ vs ($\gamma_N H_{res}$)$^{-2}$ in Fig. \ref{fig:nu_Q}. 
Based on Eq.~(\ref{eq:shift}), we estimated $^{63}\nu_Q$ from the slope of the linear line in the figure. The obtained $^{63}\nu_Q$ values, listed in Table \ref{t:ggg}, are comparable with those in other multilayered cuprates \cite{ZhengTl2223,Shimizu0234F,Julien,Tokunaga,Kotegawa2004}. 
As shown in Table \ref{t:ggg}, $^{63}\nu_Q$ decreases from 0245F($\sharp$1) to 0245F($\sharp$4); this reduction of $^{63}\nu_Q$ shows that $p$ decreases from 0245F($\sharp$1) to 0245F($\sharp$4), as observed in other hole-doped cuprates\cite{ShimizuP,Pennington,Yoshinari,Ohsugi,Zheng,Haase}.

\subsubsection{$^{63}$Cu-NMR shift and estimation of hole density $p$}

\begin{figure}[htpb]
\begin{center}
\includegraphics[width=0.8\linewidth]{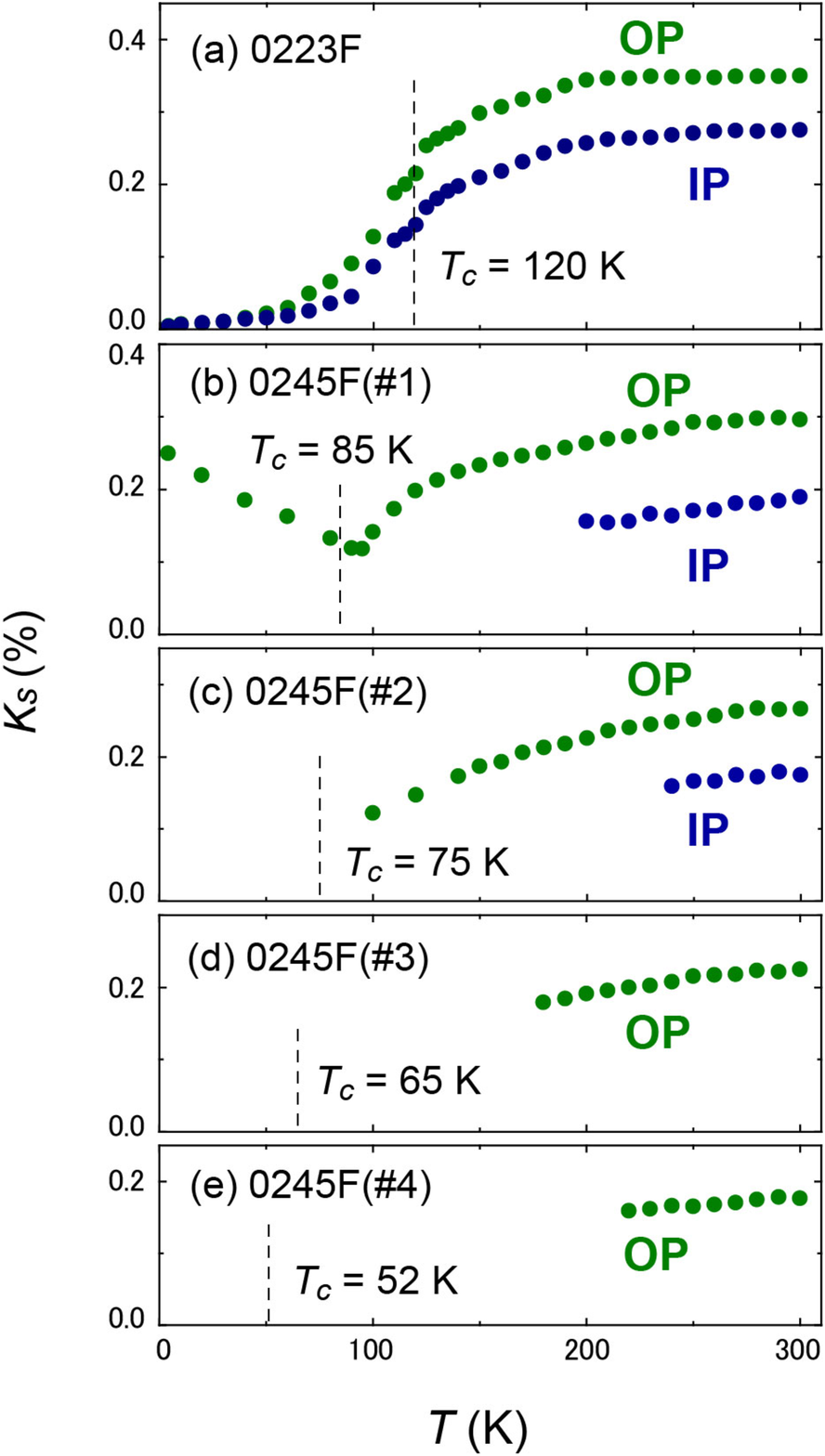}
\end{center}
\caption{\footnotesize (color online)  $T$-dependences of $^{63}$Cu Knight shift $K_s(T)$ with $H_{ex}\perp$ $c$ for (a) 0223F (cited from Ref.~\cite{ShimizuP}), (b) 0245F($\sharp$1), (c) 0245F($\sharp$2), (d) 0245F($\sharp$3), and (e) 0245F($\sharp$4). 
}
\label{fig:Ks}
\end{figure}

According to Eq.~(\ref{eq:shift}), the Knight shift $K$ for the Cu-NMR spectra in Fig. \ref{fig:NMR} is estimated by subtracting the second order quadrupole shift from the total NMR shift. In order to estimate $K$, we use the $^{63}\nu_Q$ values listed in Table~\ref{t:ggg}. 
Here, $^{63}\nu_Q$(IP) $\sim$ 7.0 to 7.5 MHz is assumed for 0245F($\sharp$3) and 0245F($\sharp$4) since $^{63}\nu_Q$ becomes smaller with decreasing $p$.
In high-$T_c$ cuprates, $K$ comprises a $T$-dependent spin part $K_s(T)$ and a $T$-independent orbital part $K_{orb}$ as follows:
\begin{equation}
K= K_{s}(T)+K_{orb}. 
\label{eq:K}
\end{equation}
The $T$-dependences of $K_s(T)$ with $H_{ex}\perp c$ for  0245F($\sharp$1), 0245F($\sharp$2), 0245F($\sharp$3), and 0245F($\sharp$4) are displayed in Figs.~\ref{fig:Ks}(b), \ref{fig:Ks}(c), \ref{fig:Ks}(d), and \ref{fig:Ks}(e), respectively. Here, $K_{orb}$ was determined as $\sim$ 0.22 \%, assuming $K_{s}(T)\simeq 0$ at a $T=$ 0 limit. 

\begin{figure}[htpb]
\begin{center}
\includegraphics[width=0.8\linewidth]{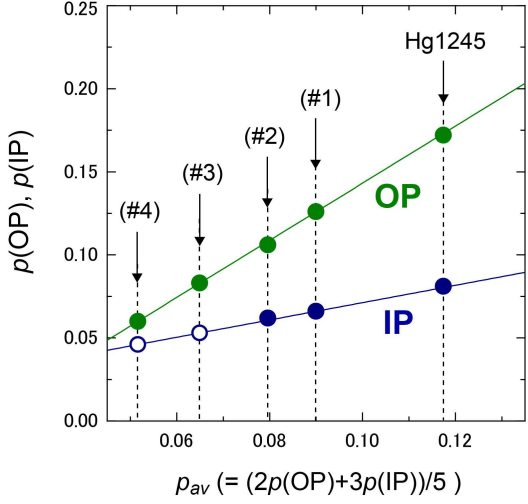}
\end{center}
\caption{\footnotesize (color online)  
Plot of $p$(OP) and $p$(IP) as function of $p_{av}$ for 0245F($\sharp$1), 0245F($\sharp$2), 0245F($\sharp$3), and 0245F($\sharp$4). Here, $p_{av}$ is the average value of $p$(OP) and $p$(IP), defined as $p_{av}$=(2$p$(OP)+3$p$(IP))/5. The data for Hg1245 is cited from Ref.~\cite{Mukuda}.
As for IP in 0245F($\sharp$3) and ($\sharp$4), it is impossible to directly estimate $p$(IP) from $K_s$(300 K). Therefore, we tentatively estimate $p$(IP) from the linear lines in the figure. As shown in the figure, $p$(OP) and $p$(IP) systematically decrease with the reduction of $p_{av}$, which allows us to extrapolate $p_{av}$ for 0245F($\sharp$3) and 0245F($\sharp$4) from $p$(OP). As for 0245F($\sharp$3), $p_{av}$=0.065 is expected from $p$(OP)=0.083 on the linear line for $p$(OP) vs $p_{av}$. Furthermore, the plot for $p$(IP) vs $p_{av}$ gives a tentative value of $p$(IP)=0.053 from $p_{av}$=0.065 as shown by an open circle. The value of $p$(IP) for 0245F($\sharp$4) is also obtained by adopting the same procedure. As summarized in Table I, $p$(OP) and $p$(IP) decrease with the increase of the fluorine content $y$, which reduces $T_c$ systematically.
}
   
\label{fig:pav}
\end{figure}

As shown in the figures, the room temperature value of $K_s(T)$ decreases with decreasing $p$ from 0245F($\sharp$1) to 0245F($\sharp$4). The values of $p$(OP) and $p$(IP), which are summarized in Table \ref{t:ggg}, are separately evaluated by using the relationship between $K_s$(300 K) and $p$ \cite{ShimizuP}. 
The quantity $p$(IP) is smaller than $p$(OP) because IP is far from charge reservoir layers, which has been usually observed in multilayered cuprates \cite{Trokiner,Julien,Kotegawa2001}.  
As for IP in 0245F($\sharp$3) and 0245F($\sharp$4), it is impossible to directly estimate $p$(IP) from $K_s$(300 K). Therefore, we tentatively estimate $p$(IP) from Fig.~\ref{fig:pav}. Figure \ref{fig:pav} shows $p$(OP) and $p$(IP) as functions of the average hole density $p_{av}$ for 0245F and another five-layered compound Hg1245 \cite{Mukuda}. Here, $p_{av}$ is defined as $p_{av}$=(2$p$(OP)+3$p$(IP))/5.
As shown in the figure, $p$(OP) and $p$(IP) systematically decrease with the reduction of $p_{av}$, which allows us to extrapolate $p_{av}$ for 0245F($\sharp$3) and 0245F($\sharp$4) from $p$(OP). As for 0245F($\sharp$3), $p_{av}$=0.065 is expected from $p$(OP)=0.083 on the linear line for $p$(OP) vs $p_{av}$. Furthermore, the plot for $p$(IP) vs $p_{av}$ gives a tentative value of $p$(IP)=0.053 from $p_{av}$=0.065. The value of $p$(IP) for 0245F($\sharp$4) is also obtained by adopting the same procedure. 
As summarized in Table \ref{t:ggg}, $p$(OP) and $p$(IP) decrease with the increase of the fluorine content $y$, which reduces $T_c$ systematically.

The $T$-dependence of $K_s(T)$ for the four 0245F samples is different from that for paramagnetic superconductors, suggesting a possible AFM order at low temperatures.   
As an example of a paramagnetic multilayered compound, we show in Fig.~\ref{fig:Ks}(a) the $T$-dependence of $K_s(T)$ for optimally-doped three-layered Ba$_2$Ca$_2$Cu$_3$O$_6$(F,O)$_2$ (0223F) with $T_c$=120 K  \cite{ShimizuP,Shimizu0223F}. 
As shown in Fig.~\ref{fig:Ks}(a), $K_s(T)$ decreases upon cooling down to $T_c$ in association with the opening of pseudogap \cite{Yasuoka,REbook}. The steep decrease of $K_s(T)$ below $T_c$ is evidence of the reduction in spin susceptibility due to
the formation of spin-singlet Cooper pairing. These behaviors in $K_s(T)$ are common in hole-doped cuprates that are paramagnetic superconductors.
On the other hand, in the case of 0245F($\sharp$1), $K_s(T)$ in Fig.~\ref{fig:Ks}(b) shows a $T$-dependence totally different from that in Fig.~\ref{fig:Ks}(a); $K_s(T)$ for IP can not be determined below $T$ $\sim$ 200 K because of the disappearance of Cu-NMR spectra, and $K_s(T)$ for OP shows an upturn at $T$ $\sim$ 85 to 90 K. These unusual behaviors of $K_s(T)$ suggest a possible AFM transition at IP in 0245F($\sharp$1) with the N\'eel temperature $T_N$ $\sim$ 85 K. 
A similar $T$-dependence in $K_s(T)$ has been reported in optimally-doped five-layered Hg1245 \cite{Mukuda,Kotegawa2004}, where $K_s(T)$ for OP shows an upturn at $T$ $\sim$ 55 K with the AFM transition at IP.
In 0245F($\sharp$2), the values of $K_s(T)$ are unmeasurable for both OP and IP at low temperatures as shown in Fig.~\ref{fig:Ks}(c). In 0245F($\sharp$3) and 0245F($\sharp$4), $K_s(T)$ for OP determined only at high temperatures, and $K_s(T)$ for IP is not obtained in all measured $T$-ranges, as shown in Figs.~\ref{fig:Ks}(d) and \ref{fig:Ks}(e).
These unusual behaviors of $K_s(T)$ suggest that AFM orders occur at both OP and IP in 0245F($\sharp$2), 0245F($\sharp$3) and 0245F($\sharp$4), and that $T_N$ increases with decreasing $p$ from 0245F($\sharp$2) to 0245F($\sharp$4).

\subsection{Evidence for AFM order probed by Cu-NQR and zero-field Cu-NMR}
\begin{figure}[htpb]
\begin{center}
\includegraphics[width=0.85\linewidth]{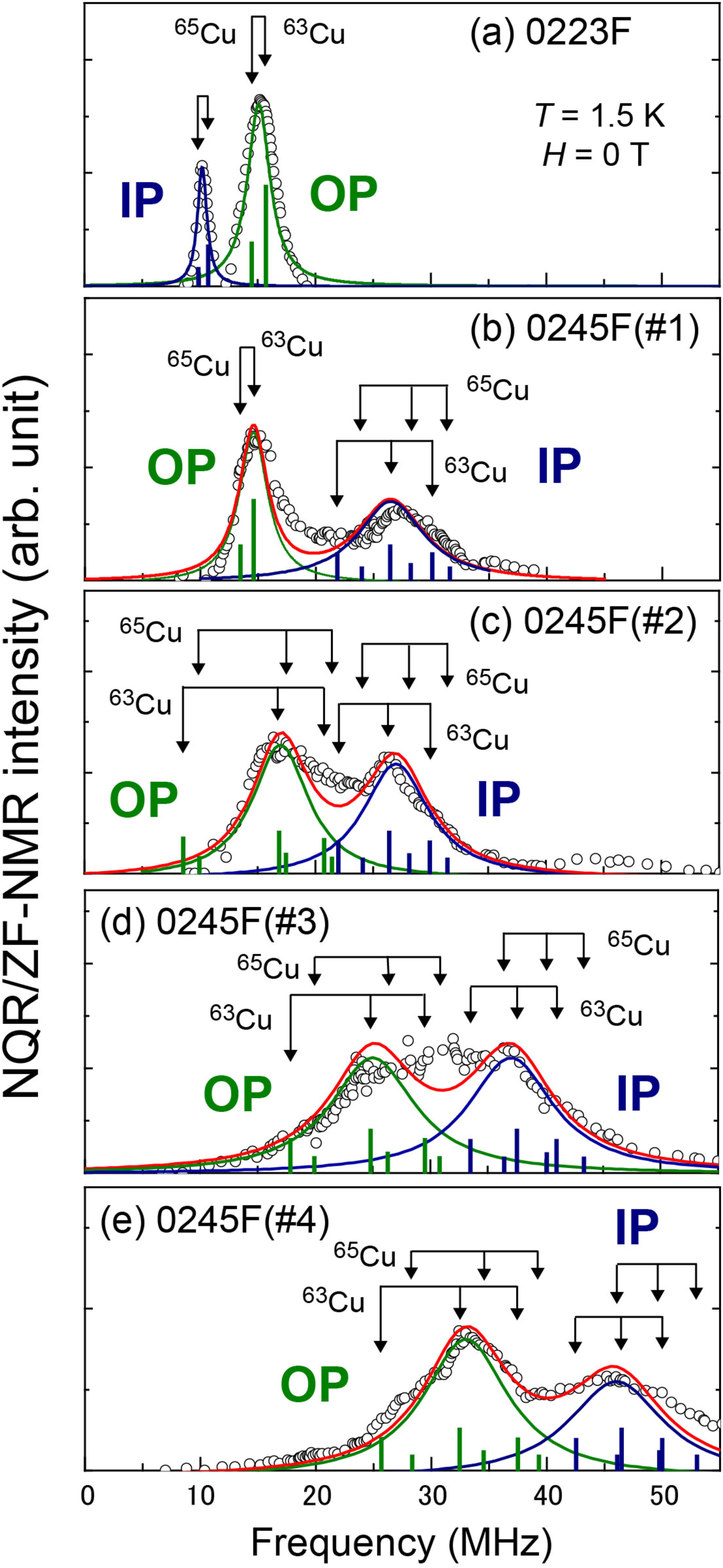}
\end{center}
\caption{\footnotesize (color online) (a) Cu-NQR spectrum at $T$=1.5 K for three-layered 0223F with $T_c$=120 K (cited from Ref. \cite{ShimizuFIN}). The spectrum is typical of paramagnetic multilayered compounds.
(b)-(e) Cu-NQR or zero-field NMR spectra at $T$=1.5 K for 0245F. The data for 0245F($\sharp$4) in (e) is cited from Ref.~\cite{ShimizuFIN}. The bars represent resonance frequencies estimated by using Eq.~(\ref{eq:hamiltonian}) with $\nu_Q$ values in Table \ref{t:ggg} and $H_{int}$ values in Table \ref{t:ggg2}. 
The curves in the figure  are Cu-NMR spectral simulations for OP and IP  based on the positions of the bars, which represents resonance frequencies estimated by using Eq. (\ref{eq:hamiltonian}), and simulations to the total Cu-NMR spectra.}
\label{fig:ZF}
\end{figure}

As shown in Fig.~\ref{fig:NMR}, the $^{63}$Cu-NMR spectra for 0245F are lost when $T$ decreases. This is due to the marked development of AFM correlations, which suggests that static AFM orders occur at low temperatures. NMR measurements at $H_{ex}$=0 T sensitively detect evidence of static AFM orders, as explained below.

In general, the Hamiltonian for Cu nuclear spins ($I=3/2$) with an axial symmetry is described in terms of the Zeeman interaction ${\cal H}_{Z}$ due to a magnetic field $H$, and the nuclear-quadrupole interaction ${\cal H}_{Q}$ as follows:
\begin{eqnarray}
{\cal H}&=&{\cal H}_Z+{\cal H}_Q  \notag \\
        &=&-\gamma_N \hbar {\bm I} \cdot {\bm H}+\frac{e^{2}qQ}{4I(2I-1)}(3I_{z^{\prime}}^2-I(I+1)),
\label{eq:hamiltonian}
\end{eqnarray}
where $eQ$ is the nuclear quadrupole moment, and $eq$ is the electric field gradient at a Cu nuclear site. In ${\cal H}_{Q}$, the nuclear quadrupole resonance (NQR) frequency is defined as $\nu_{Q}=e^{2}qQ/2h$. In paramagnetic substances, an NQR spectrum is observed due to the second term in Eq.~(\ref{eq:hamiltonian}) when $H$=$H_{ex}$=0 T. On the other hand, in magnetically ordered substances, an internal magnetic field $H_{int}$ is induced at Cu sites; in addition to the second term, the first term in Eq. (\ref{eq:hamiltonian}) contributes to the nuclear Hamiltonian even if $H_{ex}$=0 T. 
Therefore, the onset of a magnetically ordered state is observed as a distinct change of the spectral shape at $H_{ex}$=0 T.

Figure \ref{fig:ZF}(a) shows the Cu-NQR spectrum of three-layered 0223F with $T_c$=120 K \cite{ShimizuFIN} as an example of paramagnetic multilayered cuprates. The two peaks correspond to OP and IP, and both peaks include two components, $^{63}$Cu and $^{65}$Cu. It is assured that 0223F is a paramagnetic superconductor because an NQR spectrum is obtained for OP and IP at $H_{ex}$=0 T. Here, note that an NQR spectrum similar to that in Fig.~\ref{fig:ZF}(a) should be obtained at $H_{ex}$=0 T when a measured material is paramagnetic.
The spectra obtained at $H_{ex}$=0 T for five-layered 0245F are, however, totally different from that of paramagnetic 0223F as shown in Figs.~\ref{fig:ZF}(b)-\ref{fig:ZF}(e). 

Figure \ref{fig:ZF}(b) shows the spectrum measured at $H_{ex}$=0 T and $T$=1.5 K for 0245F($\sharp$1). An NQR spectrum is observed for OP, and the NQR frequency $^{63}\nu_Q$ $\sim$ 14 MHz corresponds approximately to the value estimated from Fig.~\ref{fig:nu_Q} (see Table \ref{t:ggg}). On the other hand, the spectrum for IP is different from the NQR spectrum for IP in 0223F; the resonance frequency of IP in 0245F($\sharp$1) is significantly larger than $^{63}\nu_Q$ $\sim$ 8.3 MHz, which was estimated from Fig.~\ref{fig:nu_Q}.
According to Eq.~(\ref{eq:hamiltonian}), resonance frequencies increase when $H_{int}$ is induced at Cu sites by the onset of AFM orders. The bars in Fig.~\ref{fig:ZF}(b) represent the resonance frequencies estimated by using Eq.~(\ref{eq:hamiltonian}) on the assumption of $H_{int}$ = 0 T for OP and of $H_{int}$ $\sim$ 2.3 T ($\perp$ $c$) for IP. This reveals that $H_{int}$ $\sim$ 2.3 T is induced at IP by spontaneous AFM moments $M_{AFM}$ due to the AFM order. 
As shown in the figure, there are $^{63}$Cu and $^{65}$Cu components, and each Cu component has one center peak and two satellite peaks when $H_{int}$ $\neq$ 0. However, due to the poor frequency resolution that is related to a weak signal to noise ratio, those signals are not well resolved. The weak resolution may be also attributed to the inhomogeneities of $\nu_Q$ and of the size or direction of $M_{AFM}$. Here, we tentatively represent spectra of the IP by a single Lorentzian as shown in Fig.~\ref{fig:ZF} (b).

Figures \ref{fig:ZF}(c), \ref{fig:ZF}(d), and \ref{fig:ZF}(e) show the spectra measured at $H_{ex}$=0 T and $T$=1.5 K for 0245F($\sharp$2), 0245F($\sharp$3), and 0245F($\sharp$4), respectively. In the three samples, the spectra for both OP and IP are totally different from the NQR spectra for 0223F shown in Fig.~\ref{fig:ZF}(a), which suggests AFM orders at both OP and IP.
With decreasing $p$ from 0245F($\sharp$2) to 0245F($\sharp$4), the spectrum shifts to higher frequency regions, as shown in the figures. As in the case of IP in 0245F($\sharp$1), the bars in the figures represent the resonance frequencies estimated by using Eq.~(\ref{eq:hamiltonian}). Here, we used the values of $^{63}\nu_Q$ listed in Table \ref{t:ggg} and assumed the values of $H_{int}$ listed in Table \ref{t:ggg2}.
The quantity $H_{int}$ is converted into $M_{AFM}$, as listed in Table \ref{t:ggg2}, by using the relation of $H_{int}=|A_{hf}|M_{AFM}=|A-4B|M_{AFM}$. Here, $A$ and $B$ are the on-site hyperfine field and the supertransferred hyperfine field, respectively. $A$ $\sim$ 3.7 and $B$ $\sim$ 6.1 T/$\mu_{B}$ are assumed, which are typical values in multilayered cuprates in underdoped regions \cite{ShimizuP}.
Here, we comment on some distributions of each resonance frequency presented in Fig. \ref{fig:ZF}. As mentioned above, the weak resolution may be due to the inhomogeneities of $\nu_Q$ and $H_{int}$. However, a possibility of spin-glass states is ruled out in these compounds. It is reported that the spin-glass phases cause $H_{int}$ to largely distribute in association with random directions of frozen magnetic moments, which does not allow us to obtain zero-field spectra in limited frequency regions. In fact, $^{19}$F-NMR studies in the following section assure onsets of  long-range three dimensional AFM orders.

\begin{table}[htpb]
\caption[]{List of $H_{int}$, $M_{AFM}$, and $T_N$ for 0245F. We used $H_{int}$ values presented below in order to obtain the bars in Fig. \ref{fig:ZF}. The values of $M_{AFM}$ are deduced by using the relation of $H_{int}=|A_{hf}|M_{AFM}=|A-4B|M_{AFM}$, where $A$ and $B$ are the on-site hyperfine field and the supertransferred hyperfine field, respectively. $A$ $\sim$ 3.7 and $B$ $\sim$ 6.1 T/$\mu_B$ are assumed, which are typical values in multilayered cuprates in underdoped regions \cite{ShimizuP}. The values of $T_N$ are estimated from the upturn in $K_s$(T) for 0245F($\sharp$1) and from F-NMR measurements for 0245F($\sharp$2), 0245F($\sharp$3), and 0245F($\sharp$4) (see text).}
\begin{center}
{\tabcolsep = 1.3mm
 \renewcommand\arraystretch{1.6}
  \begin{tabular}{c|ccccc}
    \hline\hline
          & $H_{int}$(OP) & $H_{int}$(IP) & $M_{AFM}$(OP) & $M_{AFM}$(IP) & $T_N$ \\
    \hline 
  $\sharp$1 &   0 T         &  2.3 T        &   --          &    0.11 $\mu_B$  &   85 K  \\
  $\sharp$2 &   1.3 T       &  2.3 T        & 0.06 $\mu_B$  &    0.11 $\mu_B$  &   95 K  \\
  $\sharp$3 &   2.1 T       &  3.3 T        & 0.10 $\mu_B$  &    0.16 $\mu_B$  &  145 K  \\
  $\sharp$4 &   2.8 T       &  4.1 T        & 0.14 $\mu_B$  &    0.20 $\mu_B$  &  175 K  \\
    \hline\hline
    \end{tabular}}
 \end{center}
\label{t:ggg2}
\end{table}

\subsection{F-NMR}

\subsubsection{$^{19}$F-NMR spectra}

\begin{figure*}[t]
\begin{center}
\includegraphics[width=0.9\linewidth]{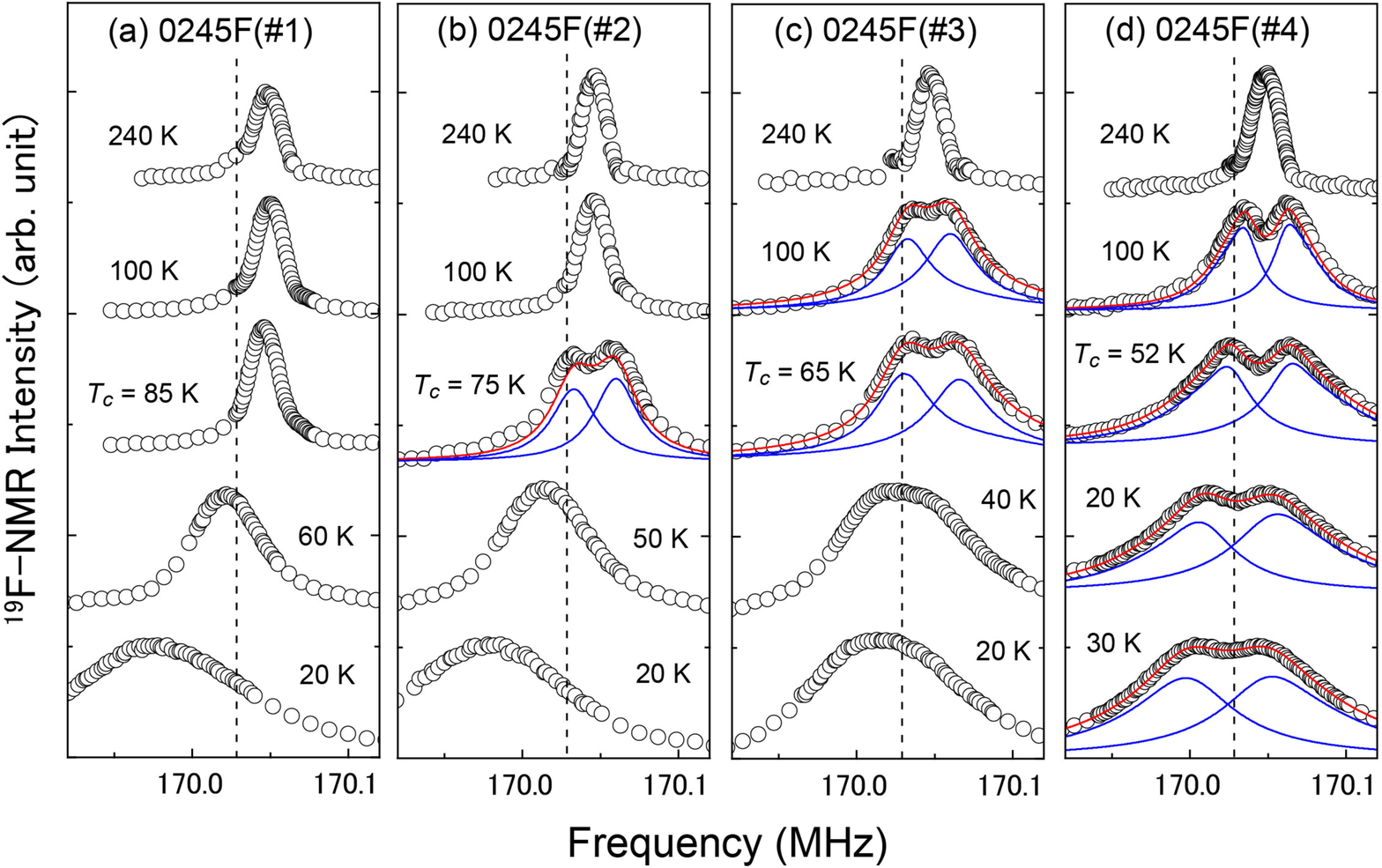}
\end{center}
\caption{\footnotesize (color online) $T$ dependences of $^{19}$F-NMR spectra for (a) 0245F($\sharp$1), (b) 0245($\sharp$2), (c) 0245F($\sharp$3), and (d) 0245F($\sharp$4). Here, $H_{ex}$ is parallel to the $c$ axis and is fixed at 4.245 T. The spectra in (b)-(d) split into two at low temperatures, pointing to the onset of AFM orders. Note that the spectrum for (a) 0245F($\sharp$1) exhibits a single peak in all measured $T$ ranges because OP is paramagnetic (see text). The dashed line indicates $^{19}K_c$=0, and the solid lines are spectral simulations to estimate $\omega$, $|H_{int,c}({\rm F})|$, and $^{19}K_c$.}
\label{fig:F-NMR}
\end{figure*}
It is difficult to deduce the $T$ dependences of $M_{AFM}$(OP) and $M_{AFM}$(IP) from the zero-field Cu-NMR measurements because the nuclear spin relaxation at Cu sites is enhanced greatly as $T$ approaches $T_N$. Instead, $^{19}$F-NMR is used to probe the $T$ dependence of the internal field ${\bm H_{int}}$(F) at apical-F sites, which is induced by $M_{AFM}$. Figure \ref{fig:F-NMR} shows the $T$ dependences of $^{19}$F-NMR spectra obtained by sweeping frequencies with $H_{ex}$ (=4.245 T) parallel to the $c$-axis.

\begin{figure*}[t]
\begin{center}
\includegraphics[width=0.85\linewidth]{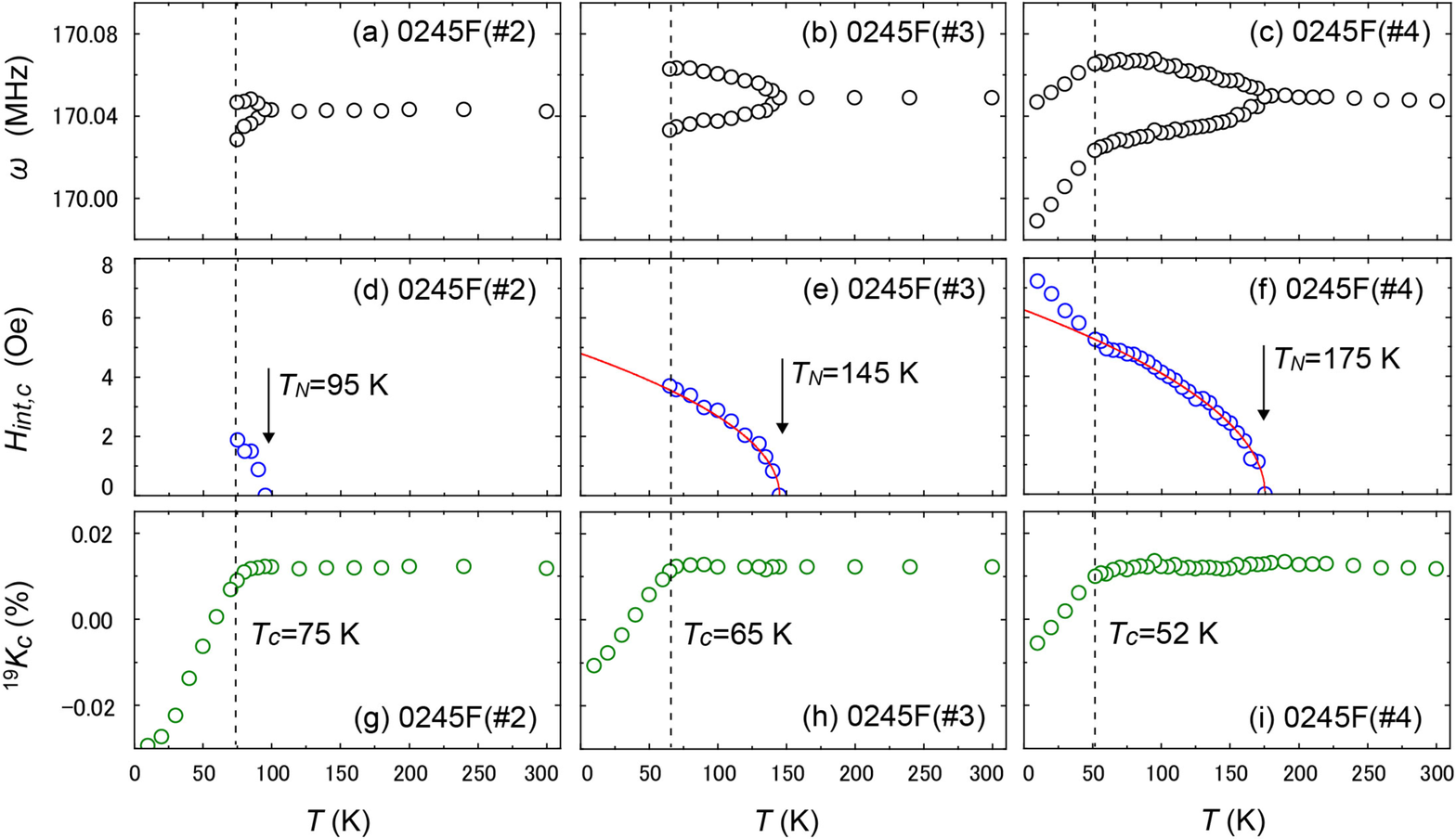}
\end{center}
\caption{\footnotesize (color online)  (a)-(c) $T$ dependence of $\omega$ for 0245F($\sharp$2), 0245F($\sharp$3), and 0245F($\sharp$4). The values of $\omega$ are estimated from spectral simulations in Fig.~\ref{fig:F-NMR}. For (a) 0245F($\sharp$2) and (b) 0245F($\sharp$3), it is impossible to determine $\omega$ below $T_c$ owing to the marked spectral broadening, which is in association with the distribution of vortices in the SC mixed state. (d)-(f) $T$ dependence of $|H_{int,c}({\rm F})|$ estimated from $\Delta \omega$=2$\times^{19}\gamma_N |H_{int,c}({\rm F})|$. Here, $\Delta \omega$ is the width of the spectral splitting in F-NMR spectra. The solid line is $H_{int,c}$(F)$\propto M_{AFM}(T)=M_{AFM}(0)(1-T/T_N)^{0.5}$, which is in good agreement with the experiment down to $T_c$. (g)-(i) $T$ dependence of Knight shift $^{19}K_c$. The decrease in $^{19}K_c$ below $T_c$ is associated with SC diamagnetic shifts.
The data in (c), (f), and (i) for 0245F($\sharp$4) are cited from Ref.~\cite{ShimizuFIN}.}
\label{fig:Fdata}
\end{figure*}

Figure \ref{fig:F-NMR}(d) shows the $T$ dependence of the F-NMR spectrum for 0245F($\sharp$4), which has the lowest $T_c$ of the present samples.
A sharp spectrum is observed with a single peak at $T$ = 240 K, but the spectrum splits into two peaks at low temperatures. As discussed before, 0245F($\sharp$4) shows the AFM order with $M_{AFM}$(OP) $\sim$ 0.14 and $M_{AFM}$(IP) $\sim$ 0.20 at $T$=1.5 K. Therefore, the spectral splitting suggests the development of ${\bm H_{int}}$(F) induced by $M_{AFM}$.
Figure \ref{fig:Fdata}(c) presents the $T$ dependence of the resonance frequency $\omega$ in the $^{19}$F-NMR spectra, which is deduced through spectral simulations shown by solid lines in Fig.~\ref{fig:F-NMR}(d). As shown in Fig.~\ref{fig:Fdata}(c), the F-NMR spectra split below $T$=175 K, suggesting the AFM ordering below the N\'eel temperature $T_N$=175 K.

Figures \ref{fig:F-NMR}(b) and \ref{fig:F-NMR}(c) show the $T$-dependences of the F-NMR spectra for 0245F($\sharp$2) and 0245F($\sharp$3), respectively.
As in the case of 0245F($\sharp$4), the F-NMR spectra split into two at low temperatures. Below $T_c$=75 K in 0245F($\sharp$2) and $T_c$=65 K in 0245F($\sharp$3), however, the two spectral peaks are blurred due to the spectral broadening related to vortex states in the SC phase.
Note that $M_{AFM}$ in 0245F($\sharp$2) and 0245F($\sharp$3) is smaller than that in 0245F($\sharp$4), and that the vortex-related spectral broadening in 0245F($\sharp$2) and 0245F($\sharp$3) is larger than that in 0245F($\sharp$4) due to the higher $T_c$ values. Therefore, in contrast to 0245F($\sharp$4), it is difficult to separate the two peaks in 0245($\sharp$2) and 0245F($\sharp$3) below $T_c$.
The $T$-dependences of $\omega$ for 0245F($\sharp$2) and 0245F($\sharp$3) are shown in Figs.~\ref{fig:Fdata}(a) and \ref{fig:Fdata}(b), suggesting the AFM transitions at $T_N$ $\sim$ 95 K and $\sim$ 145 K, respectively.

Figure \ref{fig:F-NMR}(a) shows the $T$-dependence of the F-NMR spectrum in 0245F($\sharp$1), which has the highest $T_c$ in the present samples. The spectral shape is unchanged upon cooling above $T_c$=85 K; the spectrum shifts to lower frequency regions due to the SC transition below $T_c$. As shown in Fig.~\ref{fig:ZF}(b) and listed in Table~\ref{t:ggg2}, OP in 0245F($\sharp$1) is paramagnetic at $T$=1.5 K, whereas IP shows an AFM order with $M_{AFM}$(IP) $\sim$ 0.11 $\mu_B$. Therefore, the fact that the spectrum does not split at any temperature shows that the F-NMR spectra with $H_{ex}$ parallel to the $c$ axis is not affected by $M_{AFM}$(IP), as we already discussed in the previous report \cite{ShimizuFIN}.

\subsubsection{$T$ dependence of internal field at apical-F site}

The resonance frequency $\omega$ of $^{19}$F-NMR with $H_{ex}$ parallel to the $c$-axis is expressed by  
\begin{eqnarray}
\omega \simeq {^{19}\gamma_N} H_{ex}~(1+{^{19}K_c})\pm {^{19}\gamma_N}|H_{int,c}({\rm F})|,
\label{eq:Hint}
\end{eqnarray}
where $H_{int,c}$(F) is the component of ${\bm H_{int}}$(F) along the $c$-axis, $^{19}\gamma_{N}$ is the $^{19}$F nuclear gyromagnetic ratio, and $^{19}K_c$ is the Knight shift.  The plus (minus) sign of $|H_{int,c}({\rm F})|$ in Eq.~(\ref{eq:Hint}) corresponds to its parallel (antiparallel) component along the $c$-axis.
As shown in Figs.~\ref{fig:Fdata}(a)-\ref{fig:Fdata}(c), the spectral splitting $\Delta \omega$ increases below $T_N$ due to the development of $H_{int,c}({\rm F})$, which is in proportion to $M_{AFM}$(OP). 
If the direction of $M_{AFM}$(OP) were in the basal CuO$_2$ planes with a wave vector $Q$ = ($\pi$,$\pi$), $H_{int,c}$(F) would be cancelled out at the apical-F site; the F ions are located at a magnetically symmetric position.
We have reported that $H_{int,c}({\rm F})$ is induced because $M_{AFM}$(OP) is directed out of the planes within a few degrees of the canting angle \cite{ShimizuFIN}. Refer to Ref.~\cite{ShimizuFIN} for more detailed discussions on the origin of $|H_{int,c}({\rm F})|$.

Figure \ref{fig:Fdata}(f) shows the $T$-dependence of $|H_{int,c}({\rm F})|$ for 0245F($\sharp$4), which is obtained from $\Delta \omega$=2$\times^{19}\gamma_N |H_{int,c}({\rm F})|$.
In order to focus on the $T$ evolution of $M_{AFM}$, $M_{AFM}(T)$, the relationship $H_{int,c}$(F) $\propto$ $M_{AFM}(T)$ = $M_{AFM}(0)(1-T/T_N)^{0.5}$ is displayed as the solid line in Fig.~\ref{fig:Fdata}(f). This power-law variation of $M_{AFM}(T)$ is in good agreement with the experiment down to $T_c$ = 52 K, which suggests the three-dimensional long-range order of $M_{AFM}$. 
The same $T$-dependence of $M_{AFM}(T)$ has been reported in slightly-doped LSCO compounds which exhibit AFM ground states~\cite{Borsa}.
It is also possible to reproduce the $T$ dependence down to $T_c$=52 K by another relationship $H_{int,c}$(F) $\propto$ $M_{AFM}(T)$ = $M_{AFM}(0)(1-(T/T_N)^{3/2})^{0.5}$ (not shown), which has been theoretically predicted on an itinerant AFM metal \cite{Nakayama}. In any case, the $T$ evolution of $M_{AFM}$ assures that three dimensional AFM orders set in below $T_N$.
Figure \ref{fig:Fdata}(e) shows the $T$-dependence of $|H_{int,c}({\rm F})|$ for 0245F($\sharp$3). As shown in the figure, $|H_{int,c}({\rm F})|$ evolves upon cooling following the solid line $M_{AFM}(T)$ = $M_{AFM}(0)(1-T/T_N)^{0.5}$. Below $T_c$=65 K, however, it is difficult to determine $|H_{int,c}({\rm F})|$ because the spectral peaks are blurred, as shown in Fig.~\ref{fig:F-NMR}(c). The same situation is also true in 0245F($\sharp$2); $|H_{int,c}({\rm F})|$ develops below $T_N$=95 K as shown in Fig.~\ref{fig:Fdata}(d), but the F-NMR spectral peaks are blurred below $T_c$=75 K. 
Here, note that the $T_N$ values are expected to be the same in OP and IP because three dimensional AFM interactions determine $T_N$. In other words, we expect that $M_{AFM}$(OP) and $M_{AFM}$(IP) show the same $T$ dependence, although F-NMR spectra probe the development of the staggered moment at OP below $T_N$.

\subsubsection{$^{19}$F-NMR shift evidence for SC transition in AFM background}

Here, we deal with the SC properties in 0245F. The $T$ dependence of $^{19}K_c$ is displayed in Figs.~\ref{fig:Fdata}(g)-\ref{fig:Fdata}(i). The values of $^{19}K_c$ are estimated from the gravity center of the $^{19}$F-NMR spectra.  
In all samples, $^{19}K_c$ is $T$-independent upon cooling down to $T_c$, which is different from the Knight shift $K_s(T)$ of $^{63}$Cu shown in Fig.~\ref{fig:Ks}. The reason that $^{19}K_c$ is $T$-independent above $T_c$ is that the spin component in $^{19}K_c$ is small owing to the small hyperfine coupling between $^{19}$F nuclei and Cu-$3d$ spins as reported in the literature~\cite{Kambe}. 
When decreasing $T$ below $T_c$, $^{19}K_c$ markedly decreases due to the appearance of SC diamagnetism. Here, note that the reduction of $^{19}K_c$, which is in association with the onset of high-$T_c$ SC, takes place under the background of the AFM order; these observations provide firm evidence for the uniform coexisting state of AFM and SC at a microscopic level. 

In 0245F($\sharp$4), $|H_{int,c}({\rm F})|$ additionally increases below $T_c$ as shown in Fig.~\ref{fig:Fdata}(f).
It is likely that the onset of high-$T_c$ SC decreases the size of $M_{AFM}$(OP) due to the formation of coherent spin-singlet states over the sample. Therefore, the additional increase in $|H_{int,c}({\rm F})|$ may not be attributed to an increase of $M_{AFM}$(OP) but an increase of the out-of-plane canting angle in the AFM-SC mixed state. In any case, this finding demonstrates an intimate coupling between the SC order parameter and $M_{AFM}$ \cite{ShimizuFIN}. The same behavior is expected for 0245F($\sharp$2) and 0245F($\sharp$3) as well, although the F-NMR spectral peaks are blurred due to the spectral broadening in vortex states below $T_c$.

\subsection{Phase diagram of AFM and SC in five-layered Ba$_2$Ca$_4$Cu$_5$O$_{10}$(F,O)$_2$}

\begin{figure}[htpb]
\begin{center}
\includegraphics[width=0.95\linewidth]{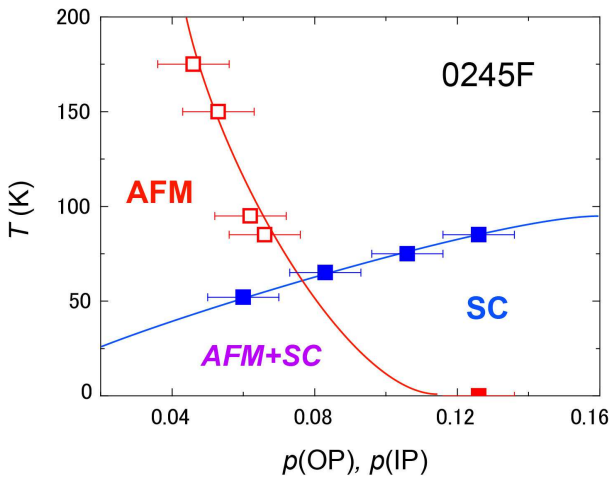}
\end{center}
\caption{\footnotesize (color online)  AFM and SC phase diagram of five-layered 0245F. 
Solid and open squares correspond to OP and IP, respectively. The values of $p$ are determined from the room $T$ values of Cu-NMR Knight shift in Fig. \ref{fig:Ks}, and those of $T_N$ are from the $T$-dependence of $|H_{int,c}({\rm F})|$ in Fig.~\ref{fig:Fdata} ( see text for detail). The solid lines in the figure are guides to the eye.
}
\label{fig:PD}
\end{figure}

Figure \ref{fig:PD} shows the phase diagram of AFM and SC in five-layered 0245F, which is derived from the present NMR study. For the SC phase, $T_c$ is plotted against $p$(OP). It has been reported that OP and IP have unique $T_c$ values, and that the higher one determines the bulk $T_c$ \cite{Tokunaga}. 
Here, it is expected in 0245F that the bulk $T_c$ listed in Table \ref{t:ggg} corresponds to $T_c$(OP). Both OP and IP are in underdoped regions, and $p$(OP) is larger than $p$(IP); therefore, $T_c$(OP) is larger than $T_c$(IP).    
For the AFM phase, $T_N$ listed in Table~\ref{t:ggg2} is plotted against $p$(IP). We assume that $T_N$ is determined by IP, considering the fact that $M_{AFM}$(IP) is larger than $M_{AFM}$(OP).
The data point for $T_N$=0 K at $p$ $\sim$ 0.126 corresponds to OP in 0245F($\sharp$1),  which is paramagnetic even at $T$=1.5 K. 

This phase diagram demonstrates that in 0245F, the AFM metallic phase is robust up to $p$ $\sim$ 0.11 to 0.12, and that the three-dimensional long-range AFM order coexists with high-$T_c$ SC in an underdoped region. Such a coexistence phase has been also reported in three-layered 0223F \cite{Shimizu0223F} and four-layered 0234F \cite{Shimizu0234F}.
Note that the coexistence of AFM and SC is not a phase separation between magnetic and paramagnetic phases; AFM and SC uniformly coexist in a CuO$_2$ plane. We have discussed the AFM ordering based on the zero-field NMR spectra shown in Fig.~\ref{fig:ZF}, where no paramagnetic signal is observed for CuO$_2$ layers in AFM states. 
Here, we determine the critical hole density $p_c$ for the AFM order as $p_c$ $\simeq$ 0.11. The value of $p_c$ is expected to exist between $p$(OP) $\sim$ 0.106 in 0245F($\sharp$2) and $p$(OP) $\sim$ 0.126 in 0245F($\sharp$1), because OP in 0245F($\sharp$1) is paramagnetic, but OP in 0245F($\sharp$2) is antiferromagnetic.

\section{Discussions}

\subsection{Layer number dependence of AFM-SC phase diagram in hole-doped cuprates}

\begin{figure}[htpb]
\begin{center}
\includegraphics[width=1.0\linewidth]{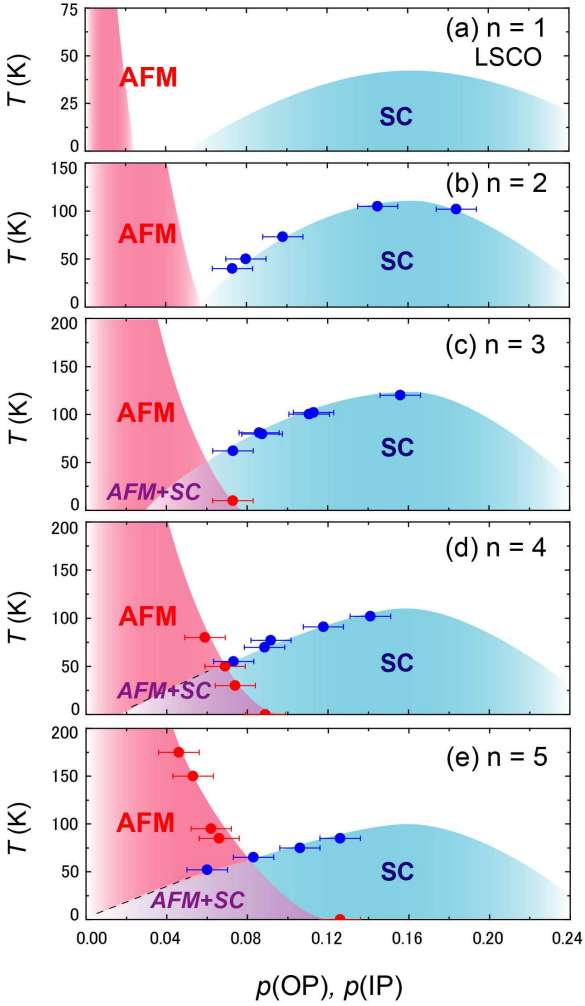}
\end{center}
\caption{\footnotesize (color online) (a) Schematic phase diagrams of LSCO. (b)-(e) AFM-SC phase diagrams of Ba$_2$Ca$_{n-1}$Cu$_n$O$_{2n}$(F,O)$_2$(02(n-1)nF). The data plot for (b) 0212F, (c) 0223F, and (d) 0234F are cited from Ref. \cite{ShimizuP,Shimizu0223F,Shimizu0234F}. The data for (e) 0245F are the same with those in Fig. \ref{fig:PD}. The $n$ dependence of the phase diagram reveals the variation of the $p_c(n)$ values: $p_c(1)\sim$ 0.02 \cite{Keimer,JulienLSCO}, $p_c(2)\sim$ 0.055 \cite{Sanna,Coneri}, $p_c(3)\sim$ 0.075 \cite{Shimizu0223F}, $p_c(4)\sim$ 0.09 \cite{Shimizu0234F,p}, and $p_c(5)\sim$ 0.11. }
\label{fig:QCP}
\end{figure}

Figure \ref{fig:QCP} shows the phase diagram of hole-doped cuprates with the different stacking number $n$ of CuO$_2$ layers. Figure \ref{fig:QCP} (a) is a schematic phase diagram for LSCO with $n$=1; figs. \ref{fig:QCP}(b)-\ref{fig:QCP}(e) show the phase diagrams of Ba$_2$Ca$_{n-1}$Cu$_n$O$_{2n}$(F,O)$_2$ (02(n-1)nF) with $n$=2 \cite{ShimizuP}, $n$=3 \cite{Shimizu0223F}, $n$=4 \cite{Shimizu0234F,p}, and $n$=5, respectively. The data in Fig.~\ref{fig:QCP}(e) correspond to those in Fig.~\ref{fig:PD}, and those in Figs.~\ref{fig:QCP}(b)- \ref{fig:QCP}(d) have been reported in previous NMR studies \cite{ShimizuP,Shimizu0223F,Shimizu0234F}. 
Here, note that the data of YBCO \cite{Sanna,Coneri} is cited as the AFM phase in Fig.~\ref{fig:QCP}(b). There are no data for the AFM phase of 0212F at present.

\begin{figure}[t]
\centering
\includegraphics[width=1.0\linewidth]{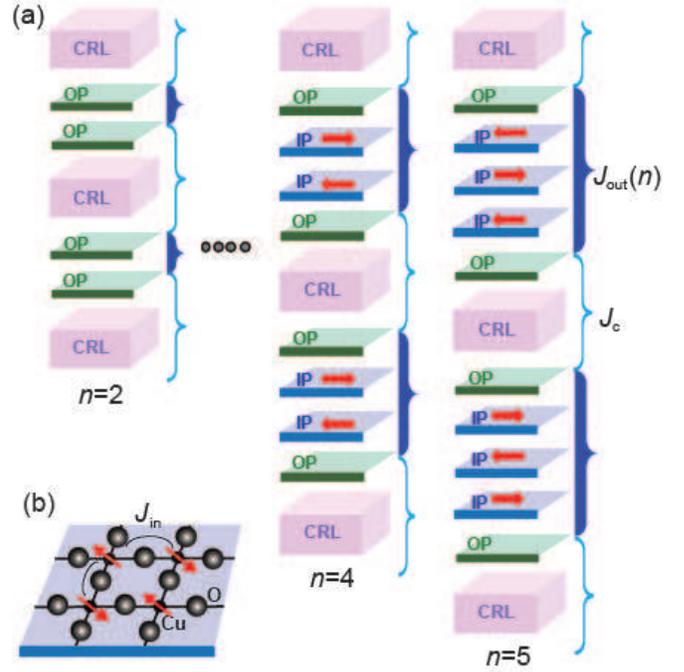}
\caption[]{\footnotesize (Color online) Schematic illustrations of magnetic couplings in $n$-layered cuprates. (a) Interlayer magnetic coupling along $c$-axis. $J_c$ is the magnetic coupling between OPs through CRL, which is independent of $n$; $J_{out}(n)$ is the magnetic coupling in a unit cell, which increases with $n$. 
(b) In-plane superexchange interaction $J_{in}$ between nearest Cu spins in two-dimensional CuO$_2$ plane. The quantity $J_{in}$ is as large as 1300~K in undoped AFM-Mott insulators \cite{JinLSCO,JinYBCO1,JinYBCO2,TokuraJ}. 
It is considered that $J_{in}$ does not depend on $n$; therefore, it is the effective interlayer magnetic coupling ($\sqrt{J_cJ_{out}(n)}$) that increases $p_c$ with increasing $n$ (see text). 
}
\label{fig:interlayer}
\end{figure}

Figures \ref{fig:QCP}(a)-\ref{fig:QCP}(e) show the variation of the $p_c$ values for $n$-layered compounds, $p_c(n)$: $p_c(1)$ $\sim$ 0.02 \cite{Keimer,JulienLSCO}, $p_c(2)$ $\sim$ 0.055 \cite{Sanna,Coneri}, $p_c(3)$ $\sim$ 0.075 \cite{Shimizu0223F}, $p_c(4)$ $\sim$ 0.09 \cite{Shimizu0234F,p}, and $p_c(5)$ $\sim$ 0.11.
This increase of $p_c(n)$ is qualitatively explained as a result of the fact that the interlayer magnetic coupling becomes stronger with increasing $n$.
The mother compounds of high-$T_c$ cuprates are characterized by a large in-plane superexchange interaction $J_{in}$ $\sim$ 1300 K between nearest-neighboring Cu-spins \cite{JinLSCO,JinYBCO1,JinYBCO2,TokuraJ}.
In addition to $J_{in}$, an interlayer magnetic coupling along the $c$-axis plays a crucial role in stabilizing an AFM order since no long-range AFM order occurs in an isolated two-dimensional system at a finite temperature.
The effective interlayer magnetic coupling of $n$-layered cuprates is given as $\sqrt{J_cJ_{out}(n)}$, where $J_c$ is a magnetic coupling between OPs through CRL (charge reservoir layer), and $J_{out}(n)$ is that in a unit cell, as illustrated in Fig. \ref{fig:interlayer}(a). 
It is considered that $J_c$ is independent of $n$, and that $J_{out}(n)$ increases with increasing $n$; therefore, the weakness of $J_{out}(n)$ suppresses the static long-range AFM orders in LSCO and YBCO at such small doping levels. 

While $p_c(n)$ increases with $n$, it seems to saturate at $\sim$ 0.14 to 0.16 even in the strong limit of interlayer magnetic coupling expected for infinite-layered compounds; $M_{AFM}$ at the ground state is extrapolated to zero at $p$ $\sim$ 0.14 to 0.16, as discussed in the next section (see Fig. \ref{fig:ZeroT}).
These results suggest that the uniform coexistence of AFM and SC is a universal phenomenon in underdoped regions, when the interlayer magnetic coupling is strong enough to stabilize an AFM ordering. A recent report seems to support this conclusion from a theoretical point of view \cite{YamasearXiv}.

\subsection{Ground-state phase diagram of CuO$_2$ plane}

\begin{figure*}[t]
 \begin{center}
  \includegraphics[width=0.8\linewidth]{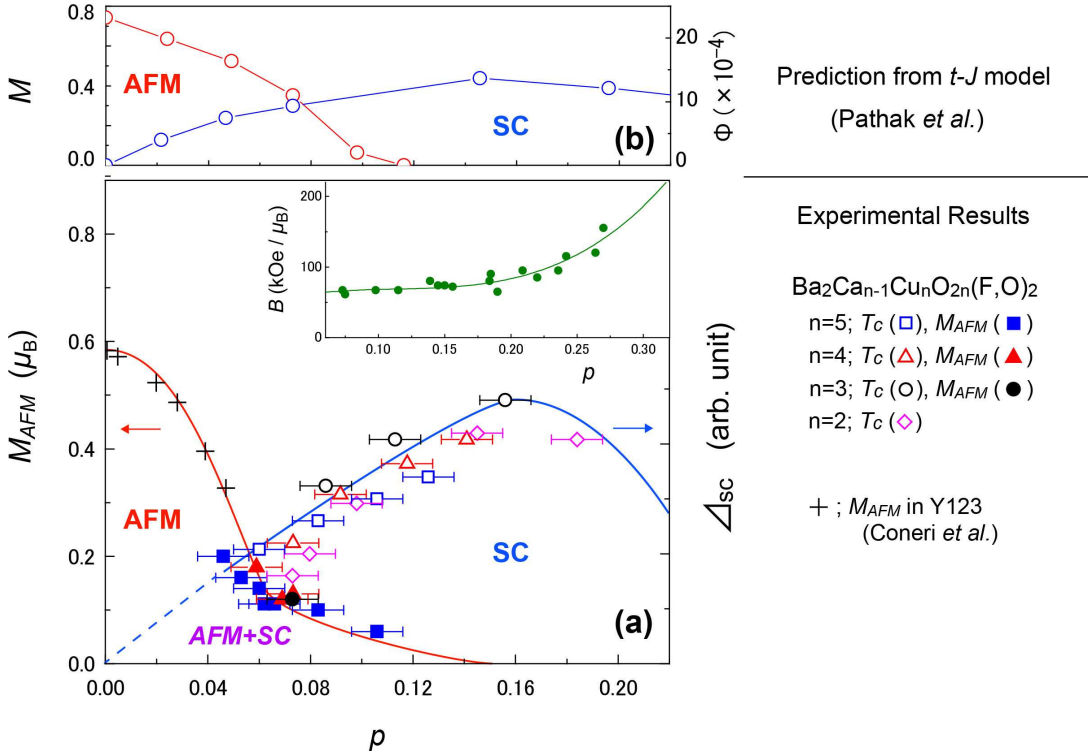}
 \end{center}
 \caption{\footnotesize (color online) (a) Ground-state phase diagram of high-$T_c$ cuprates. The quantities of $M_{AFM}$ and $\Delta_{SC}$ are plotted as a function of $p$. The data for 0212F, 0223F, and  0234F are cited from Refs. \cite{ShimizuP,Shimizu0223F,Shimizu0234F}, and the data for Y123 are cited from Ref.~\cite{Coneri}. Here, it is assumed that $\Delta_{SC}$ is proportional to $T_c$.  
The quantity $M_{AFM}$ exists up to the quantum critical hole density $p_{qcp}$ $\sim$ 0.14 to 0.16, at which $M_{AFM}$ disappears even at the ground state.
Note that the maximum of $\Delta_{SC}$ (i.e., $T_c$) is around $p$ $\sim$ 0.16, which is close to $p_{qcp}$. 
The inset shows the $p$-dependence of the supertransferred hyperfine fields $B$ at Cu sites, which originates from the hybridization between Cu($3d_{x^2-y^2}$) and O($2p\sigma$) orbits. The hybridization between Cu($3d_{x^2-y^2}$) and O($2p\sigma$) orbits drastically increases as $p$ increases above $p$ $\sim$ 0.16 to 0.18, which would correspond to the disappearance of $M_{AFM}$ and the suppression of $T_c$ (see text).
The solid lines in the figure are guides to the eye. (b) AFM and SC phase diagram predicted by the $t$-$J$ model \cite{Pathak}. $M$ and $\Phi$ are the AFM and SC order parameters, respectively.}
 
 \label{fig:ZeroT}
\end{figure*}

Figure \ref{fig:ZeroT}(a) shows $M_{AFM}$ at $T$=1.5 K and the SC energy gap $\Delta_{SC}$ as a function of $p$. The data include 0245F and the previous reports for 0212F, 0223F, and 0234F \cite{ShimizuP,Shimizu0223F,Shimizu0234F,ShimizuFIN}. 
The values of $M_{AFM}$ are estimated from zero-field NMR measurements, and those of $\Delta_{SC}$ are assumed to be proportional to $T_c$. We also plot $M_{AFM}$ for Y123 \cite{Coneri} in the figure instead of that for bi-layered 0212F since the AFM phase of 0212F has not been reported. Note that, in contrast to Fig.~\ref{fig:QCP}, this figure focuses on the ground state of CuO$_2$ planes, which gives us an opportunity to compare the experimental results with theoretical outcomes.

The AFM phase in Fig.~\ref{fig:ZeroT}(a) is characterized by the fact that $M_{AFM}$ exists up to the quantum critical hole density $p_{qcp}$ $\sim$ 0.14 to 0.16, at which $M_{AFM}$ disappears even at the ground state. 
It is worth noting that irrespective of $n$, the maximum of $\Delta_{SC}$ (i.e., $T_c$) is around $p$ $\sim$ 0.16, which is close to $p_{qcp}$ $\sim$ 0.14. This implies an intimate relationship between the SC and AFM order parameters. 
We also point out that at finite temperatures, no static long range AFM order is observed when $p_c(n) < p$, as shown in Fig. \ref{fig:QCP}, although $M_{AFM}$ is expected to exist at the ground state when $p_c(n) < p \le p_{qcp}$. 
In the $p$ range, AFM moments are fluctuating at finite temperatures, which would induce various anomalies in underdoped regions. 
As a matter of fact, spin-glass phases or stripe phases have been reported in $p_c(n) < p \le p_{qcp}$  when $n$=1 and 2 \cite{JulienLSCO,Coneri,Albenque,Alloul}; AFM moments fluctuating due to weak interlayer magnetic couplings may be frozen in association with disorders related to chemical substitutions, a buckling of CuO$_2$ planes, and  the onset of a charge order.

As for the SC phase, $\Delta_{SC}$ gradually decreases with decreasing $p$ from the optimally-doping level, $p$ $\sim$ 0.16. The size of $\Delta_{SC}$ (i.e., $T_c$) at $p$ $\sim$ 0.16 is largest when $n$=3, which is a universal trend in multilayered cuprates \cite{Scott,IyoTc}. When we extrapolate the hole density $p^{*}$ at which $\Delta_{SC}$ = 0, $p^{*}$ is $\sim$ 0.05-0.06 for $n$=2; however, $p^{*}$ probably exists at below $\sim$ 0.05 for $n$=5. This variation of $p^{*}$ is clearly seen in Fig.~\ref{fig:QCP}, where it seems that $p^{*}$ moves to more underdoped regions as $n$ increases. 
The reason for a possible reduction in $p^{*}$ is partly attributed to the flatness of the CuO$_2$ planes. With increasing $n$, it is expected that the flatness of CuO$_2$ planes is enhanced because the disorder effect related to heterovalent substitutions in CRL is relieved, especially at IP. Therefore, it would be possible that hole carriers slightly doped into Mott insulator can move even at low temperatures without Anderson localization when $n$=5, whereas they localize when $n$=1 and 2 \cite{Borsa,Coneri,IshidaAya}. Further investigations on extremely underdoped regions for multilayered compounds remain as future works.  

We propose Fig.~\ref{fig:ZeroT} as the ground state phase diagram of CuO$_2$ planes. 
The ground state in underdoped regions is characterized by the uniformly mixed state of AFM and SC, which is observed even at finite temperatures when the interlayer magnetic coupling is strong enough to stabilize AFM orders. 
Another important point is that $\Delta_{SC}$ (i.e., $T_c$) begins to decrease at $p$ $\sim$ 0.16 with increasing $p$. This is because a considerable change of electronic states occurs at $p$ $\sim$ 0.16. The inset of Fig.~\ref{fig:ZeroT} shows the $p$ dependence of the supertransferred hyperfine field $B$ at Cu sites in various hole-doped cuprates \cite{ShimizuP}; $B$ originates from the hybridization between Cu($3d_{x^2-y^2}$) and O($2p\sigma$) orbits. 
As shown in the inset, the hybridization between Cu($3d_{x^2-y^2}$) and O($2p\sigma$) orbits in overdoped regions are much larger than those in underdoped regions. This event corresponds to the suppression of the onsite Coulomb repulsion $U$ or the in-plane superexchange interaction $J_{in}$, which leads to the disappearance of $M_{AFM}$ and the suppression of $T_c$.

The ground state phase diagram presented here is qualitatively consistent with theoretically predicted ones in terms of either the $t$-$J$ model~
\cite{Chen,Giamarchi,Inaba,Anderson1,Zhang,Himeda,Kotliar,TKLee,Demler,Shih,Paramekanti,Anderson2,Ogata,Pathak}, or the Hubbard model in a strong correlation regime~\cite{Senechal,Capone}. 
Figure \ref{fig:ZeroT}(b) shows an AFM and SC phase diagram predicted by the $t$-$J$ model with the condition that the second-nearest-neighbor hopping $t^{\prime}$ is zero \cite{Pathak}. Here, values for the vertical axis, $M$ and $\Phi$, in Fig. \ref{fig:ZeroT}(b) are an AFM and a SC order parameters, respectively. See Ref. \cite{Pathak} for details on $t^{\prime}$, $M$, and $\Phi$. 
As shown in the figure, the AFM order parameter decreases with increasing $p$ and vanishes at around $p$ $\sim$ 0.12 and coexists with the SC order parameter in underdoped regions. Our experimental results demonstrate a good consistency with the $t$-$J$ model; we conclude here that the large $J_{in}$, which attracts electrons of opposite spins at neighboring sites, is the origin for high-$T_c$ SC.

\section{Conclusion}

Site-selective Cu- and F-NMR studies have unraveled the intrinsic phase diagram of AFM and SC in hole-doped cuprates Ba$_2$Ca$_{n-1}$Cu$_n$O$_{2n}$(F,O)$_2$ (02(n-1)nF). The obtained results are as follows:
\begin{enumerate}
\item[(i)] The AFM metallic state is robust up to the critical hole density $p_c$ for AFM orders, and it uniformly coexists with SC. 
\item[(ii)] The $p_c$ for $n$-layered compounds, $p_c(n)$, increases with increasing $n$ due to the growth of the interlayer magnetic coupling. The $p_c(n)$ values are deduced as $p_c(1)$ $\sim$ 0.02 \cite{Keimer,JulienLSCO},  $p_c(2)$ $\sim$ 0.055 \cite{Sanna,Coneri}, $p_c(3)$ $\sim$ 0.075 \cite{Shimizu0223F}, $p_c(4)$ $\sim$ 0.09\cite{Shimizu0234F}, and $p_c(5)$ $\sim$ 0.11.
\item[(iii)]  The maximum of $\Delta_{SC}$ (i.e., $T_c$) takes place around $p$ $\sim$ 0.16, which is just outside the quantum critical hole density $p_{qcp}$ $\sim$ 0.14 to 0.16. 
\item[(iv)]  The ground-state phase diagram of AFM and SC is in good agreement with that theoretically predicted by the $t$-$J$ model\cite{Chen,Giamarchi,Inaba,Anderson1,Zhang,Himeda,Kotliar,TKLee,Demler,Shih,Paramekanti,Anderson2,Ogata,Pathak} or by the Hubbard model in the strong correlation regime~\cite{Senechal,Capone}.
\end{enumerate}

We conclude that those results are accounted for by the {\it Mott physics} based on the $t$-$J$ model. In the physics behind high-$T_c$ phenomena, there is a very strong Coulomb repulsive interaction $U$~($>$6 eV), which  prohibits the double occupancy of an up-spin electron and a down-spin electron at the same site. 
When noting that the strength of $U$ is almost unchanged by hole doping, it is considered that the large $J_{in}$ attracts electrons of opposite spins at neighboring sites \cite{Anderson1,Ogata,Anderson2}. The qualitative consistency in the ground-state phase diagram between our experimental results and theoretical ones support the $t$-$J$ model as a mechanism of high-$T_c$ SC.

\section*{Acknowledgement}

The authors are grateful to M. Mori and T. Tohyama for his helpful discussions. This work was supported by a Grant-in-Aid for Specially promoted Research (20001004) and for Young Scientists (B) (23740268) and  by the Global COE Program (Core Research and Engineering of Advanced Materials-Interdisciplinary Education Center for Materials Science) from The Ministry of Education, Culture, Sports, Science and Technology, Japan.


\clearpage

\end{document}